\documentclass{article}
\usepackage{amsmath,amssymb,enumerate,eucal,longtable, theorem,xcolor}

\numberwithin{equation}{section}

\theoremheaderfont{\normalfont\scshape}
{\theorembodyfont{\rmfamily}}
{\theorembodyfont{\rmfamily}}

\title{Stellar flybies within 1 ly from the Sun and stars passing through the Hills cloud}
\author{Igor Yu. Potemine \\ 
\small Institut de Mathématiques, Universit\'e Paul Sabatier, Toulouse, France}
\date{}

\begin{document}
\maketitle
\abstract{We reexamine the SIMBAD database with incorporated Gaia DR3 parallaxes and proper motions. Appropriate query searches allow us to find several nearby stars with measured radial velocities having flybies within 1 ly from the Sun (in linear approximation). The closest past flyby $\approx 215.7$ kyr ago at $\approx 0.136$ pc is attributed to the star UCAC4 323-037188, currently located at 21.74 pc from the Sun. If the radial velocity of a star is unknown, we attribute to it an ``advantageous'' value of $\pm 50$ km/s. It allows us to create an additional list of ``Nemesis candidates'' for the perforation of the inner Oort cloud. The closest potential flyby at $\approx 0.015$ pc ($\approx$ 0.050 ly $\approx$ 3147 AU) from the Sun is attributed to GALEX J013712.5-012958 (Gaia DR3 2508809660245711360). It is currently located at the distance of about 111.1 pc and belongs to the Pisces-Eridanus stream. Also, close flybies of some massive bright stars (Algol, A-giant HD 107914, A-subgiant HD 2733 and B-subgiant HD 165704) are analyzed. Finally, we notice that B1-star BD+60 596, whose proper motions and radial velocity are both small, can be considered as the Sun's pseudo-stream companion.}

\medskip

\tableofcontents

\section{Introduction}

In 1916 an American astronomer E. Barnard discovered that the star 
BD$+04$ 3561a has an extraordinary high proper motion of 10.3 arcseconds per year. It is literally flying in the sky.

\smallskip There is also a different phenomenon of stars with small proper motions but high radial velocities. Their trajectories are almost parallel to the Sun's trajectory. In addition, these stars furtively approach the Sun and then quickly run away.

\smallskip Many authors (cf.~\cite{Gar99,Gar01,Dyb06,Bob10}) have searched past and future stellar perturbers of the Oort cloud. Indeed, it turns out that a massive star passing within 1 ly could have a significant influence on long-period comets. More closely, at 10000 AU, such a star would have a serious direct influence on the inner Oort cloud and trans-Neptunian objects.

\smallskip The most known example is the K7-star Gliese 710 with closest approach at $10635\pm 500$ AU in 1.29 Myr \cite{dFM22a}. In recent papers (cf.~\cite{BB, Dyb23, dFM22b}), there appeared also the Scholz's star and G3-star HD 7977.

\smallskip In the paper \cite{Pot13}, we examined an additional Nemesis candidate HD 107914 / HIP 60503 with very small proper motions. New results for this remarkable A7/8 giant are presented in section 6.

\smallskip In the present article, we establish an updated list of Nemesis candidates and Hills cloud perturbers.

\section{Minimal distances and times of closest approaches}

We use Julian years and light-years so that 1 parsec $= p\approx 3.261563777$ ly and 1 km/s corresponds to the velocity of $c^{-1}$ ly/y where $c\approx 299792.458$ is the speed of light in km/s.

\smallskip Let $X$ be a star with parallax $\pi_X^{}$ (in mas) at the current distance of $d_X = p\cdot 10^3/\pi_X^{}$ ly from the Sun denoted by the symbol $\odot$. The minimal distance $d_{\odot X}^{\mathrm{min}}$
from $X$ to $\odot$ is equal to the radius $R$ of the sphere centered
at the Sun and tangent to the trajectory of $X$. It is easy to show that
\begin{equation}
\left|\frac{v_r}{v_t}\right|=\sqrt{\left(\frac{d_X}{R}\right)^2-1}
\end{equation}
where $v_r$ and $v_t$ are radial and transverse (tangential) velocities of a star $X$ respectively. The transverse velocity can be easily calculated from available catalogue values:
\begin{equation}
v_t = \frac{\pi c}{648\cdot 10^6} \mu_T d_X =\frac{\pi c p}{648\cdot 10^3}
\frac{\mu_T^{}}{\pi_X^{}}\approx 4.74047\cdot\frac{\mu_T^{}}{\pi_X^{}}\ \mathrm{km/s}
\end{equation}
where $\mu_T$ is the total proper motion of $X$ in mas and $\frac{\pi}
{648\cdot 10^6}$ the number of radians in 1 mas. From these two formulas we obtain:
\begin{equation}
d_{\odot X}^{\mathrm{min}}=R=\frac{d_X}{\sqrt{K^2 + 1}}\ \mathrm{where}\ 
K = \frac{v_r}{v_t} \approx \frac{v_r\pi_X^{}}{4.74047\cdot\mu_T^{}}.
\end{equation}
The time of closest approach $t_{\odot X}^{\mathrm{min}}$ in years can be calculated, for example, by the formula:
\begin{equation}
t_{\odot X}^{\mathrm{min}}=\pm\frac{\sqrt{d_X^2-R^2}}{\sqrt{v_r^2+v_t^2}}
\cdot c.
\end{equation}
The sign is ``$+$'' when $v_r^{}$ is negative and vice versa. It gives also a simpler formula
\begin{equation}
t_{\odot X}^{\mathrm{min}}=-\frac{d_X^{}v_r^{}}{v_r^2+v_t^2}
\cdot c.
\end{equation}

\section{Parametric frames for SIMBAD query searches}

We will indicate here a systematic way to find candidates for flybies within 1 ly in the SIMBAD database.

\smallskip Suppose that $R\leqslant 1$ ly. From the formulas above we get
\begin{equation}
|K|\geqslant \sqrt{d_X^2-1}\ \textrm{and}\ v_t^{}\leqslant 
\frac{|v_r^{}|}{\sqrt{d_X^2-1}}.
\end{equation}
Thus, we obtain,
\begin{equation}
\mu_{\,T}^{}\lessapprox\frac{|v_r^{}|\cdot\pi_X^{}}{4.74047\cdot \sqrt{d_X^2-1}}.
\end{equation}
When $\pi_X\ll 1000$ is small enough then
\begin{equation}
\mu_{\,T}^{}\lessapprox\frac{|v_r^{}|\cdot\pi_X^2}{4740.47\cdot p}.
\end{equation}
For example, if $\pi_X^{}\approx 10$ mas and $|v_r^{}|\approx 50$ km/s then $\mu_{\,T}\lessapprox 0.324$ mas/yr.

\smallskip Using formulas above, one can create an appropriate grid of key values of $\pi_X^{}$ and $v_r^{}$ and make SIMBAD query searches.

\section{Closest stellar flybies within 1 ly}

All following ``Nemesis candidates'' can be found by a straightforward SIMBAD search. Our results are consistent with calculations in the recent paper \cite{dFM22b}.

\vskip 0.5 cm \ 

\begin{center}
\begin{longtable}{|l|l|l|l|l|}
\caption{List of Nemesis candidates.} \label{tab:nemesis} \\

\hline \multicolumn{1}{|c|}{\textbf{ID}} 
& \multicolumn{1}{c|}{$\pi_X^{}$} 
& \multicolumn{1}{c|}{prop. mot.}
& \multicolumn{1}{c|}{$v_t^{}$}
& \multicolumn{1}{c|}{$v_r^{}$} \\ 
& \multicolumn{1}{c|}{(mas)} 
& \multicolumn{1}{c|}{(mas/yr)} 
& \multicolumn{1}{c|}{(km/s)} 
& \multicolumn{1}{c|}{(km/s)} \\ \hline
\endfirsthead

\multicolumn{5}{c}%
{{\bfseries \tablename\ \thetable{} -- continued from previous page}} \\
\hline \multicolumn{1}{|c|}{\textbf{ID}} 
& \multicolumn{1}{c|}{$\pi_X^{}$}
& \multicolumn{1}{c|}{prop. mot.} 
& \multicolumn{1}{c|}{$v_t^{}$}  
& \multicolumn{1}{c|}{$v_r^{}$}\\ \hline 
\endhead

\hline \multicolumn{5}{|r|}{{Continued on next page}} \\ \hline
\endfoot

\hline
\endlastfoot

Scholz's star    &148.8   &$-$52.7  $-$95.8  &3.483 &$+$83.10 \\
GJ 710           &52.3963 &$-$0.414 $-$0.108 &0.039 &$-$14.42 \\
UCAC4 556-008431 &47.8649 &$+$0.041 $-$4.601 &0.456 &$+$31.08 \\
UCAC4 323-037188 &45.9926 &$-$3.599 $+$4.802 &0.619 &$+$98.58 \\
UCAC4 364-018255 &25.6277 &$-$0.328 $-$1.298 &0.248 &$+$45.18 \\	
HD 7977          &13.2118 &$+$0.144 $+$0.010 &0.052 &$+$26.76 \\
UCAC4 237-008148 &10.2340 &$+$0.094 $+$0.461 &0.218 &$+$82.48

\end{longtable}
\end{center}

\smallskip Close encounters are summarized in the following table. The closest past flyby is attributed to the star UCAC4 323-037188, currently located at 21.74 pc from the Sun.

\begin{center}
\begin{longtable}{|l|l|l|l|l|}
\caption{Closest stellar flybies.} \label{tab:flyby} \\

\hline \multicolumn{1}{|c|}{\textbf{ID}} 
& \multicolumn{1}{c|}{$d_{\odot X}^{\mathrm{min}}$} 
& \multicolumn{1}{c|}{$d_{\odot X}^{\mathrm{min}}$}
& \multicolumn{1}{c|}{$d_{\odot X}^{\mathrm{min}}$}
& \multicolumn{1}{c|}{$t_{\odot X}^{\mathrm{min}}$} \\ 
\multicolumn{1}{|c|}{\textbf{}} 
& \multicolumn{1}{c|}{(pc)} 
& \multicolumn{1}{c|}{(ly)}
& \multicolumn{1}{c|}{(AU)}
& \multicolumn{1}{c|}{(kyr)} \\
\hline
\endfirsthead

\multicolumn{5}{c}%
{{\bfseries \tablename\ \thetable{} -- continued from previous page}} \\
\hline \multicolumn{1}{|c|}{\textbf{ID}} 
& \multicolumn{1}{c|}{$d_{\odot X}^{\mathrm{min}}$} 
& \multicolumn{1}{c|}{$d_{\odot X}^{\mathrm{min}}$}
& \multicolumn{1}{c|}{$d_{\odot X}^{\mathrm{min}}$}
& \multicolumn{1}{c|}{$d_{\odot X}^{\mathrm{min}}$} \\ \hline
\endhead

\hline \multicolumn{5}{|r|}{{Continued on next page}} \\ \hline
\endfoot

\hline
\endlastfoot

GJ 710           &0.051 &0.167 &10568 &$+$1294.1 \\
UCAC4 323-037188 &0.136 &0.445 &28138 &$-$0215.7 \\
HD 7977          &0.146 &0.478 &30216 &$-$2765.6 \\
UCAC4 364-018255 &0.214 &0.698 &44115 &$-$0844.5 \\
UCAC4 237-008148 &0.258 &0.842 &53254 &$-$1158.4 \\
Scholz's star    &0.281 &0.918 &58054 &$-$0078.9 \\
UCAC4 556-008431 &0.306 &0.999 &63176 &$-$0657.1

\end{longtable}
\end{center}
We don't consider white dwarf candidates, like UPM J0812-3529 (cf.~\cite{BJ}), having suspiciously high radial velocities, superior to 
$\pm 200$ km/s.

\section{Extended list of Nemesis candidates}

We use parametric frames of section 3 in order to create an extended list of ``Nemesis candidates'' with unknown radial velocities, currently located within 200 pc. Stars with big measurement errors are excluded. By default, we indicate Gaia DR3 identifiers.

\begin{center}
\begin{longtable}{|l|l|l|l|}
\caption{Extended list of candidates (continuing Table 1).} \label{tab:ext} \\

\hline \multicolumn{1}{|c|}{\textbf{ID}} 
& \multicolumn{1}{c|}{$\pi_X^{}$} 
& \multicolumn{1}{c|}{proper motions}
& \multicolumn{1}{c|}{$v_t^{}$} \\ 
& \multicolumn{1}{c|}{(mas)} 
& \multicolumn{1}{c|}{(mas/yr)} 
& \multicolumn{1}{c|}{(km/s)} \\ \hline
\endfirsthead

\multicolumn{4}{c}%
{{\bfseries \tablename\ \thetable{} -- continued from previous page}} \\
\hline \multicolumn{1}{|c|}{\textbf{ID}} 
& \multicolumn{1}{c|}{$\pi_X^{}$}
& \multicolumn{1}{c|}{prop. mot.} 
& \multicolumn{1}{c|}{$v_t^{}$} \\ \hline 
\endhead

\hline \multicolumn{4}{|r|}{{Continued on next page}} \\ \hline
\endfoot

\hline
\endlastfoot

WISE J053516.80-750024.9 &250.000 &$-$127 $+$13 &2.421\\
WISE J153541.67+222525.5 &90.7878 &$-$2.375 $-$19.281 &1.014 \\
UCAC4 767-032810       &79.7311 &$+$11.736 $+$12.796 &1.032 \\
UCAC4 638-030976       &68.5664 &$+$1.334 $+$0.085 &0.092 \\
UCAC4 563-099325       &67.4697 &$-$8.371 $-$4.825 &0.679 \\
UCAC4 486-052109       &59.4668 &$+$8.667 $-$7.582 &0.918 \\
UCAC4 563-093667       &57.2635 &$-$7.200 $-$1.058 &0.602 \\
GALEX J002332.9+432030 &53.4296 &$-$2.841 $-$1.834 &0.300 \\
4198241063381323264    &49.8384 &$-$1.295 $-$5.625 &0.522 \\
5614623799042319360    &41.2974 &$-$1.283 $-$4.432 &0.530 \\
2031432910114027648    &41.0214 &$+$1.439 $-$3.870 &0.477 \\
1835040758820804608    &39.5958 &$+$1.195 $-$1.283 &0.210 \\
UCAC4 362-032250       &38.5171 &$-$0.606 $+$2.903 &0.365 \\
4143776239543076864    &36.6095 &$-$0.048 $-$3.264 &0.423 \\
5615515773536993920    &35.5428 &$-$0.693 $+$3.922 &0.531 \\
1940843120604424576    &35.1943 &$-$0.196 $+$0.425 &0.063 \\
4158060785356113792    &33.2454 &$-$2.177 $-$1.184 &0.353 \\
5614676330771453440    &33.2324 &$+$0.437 $-$1.870 &0.274 \\
2027191994999462016    &31.8573 &$+$0.590 $-$2.996 &0.454 \\
4256872795844365568    &31.6929 &$-$0.575 $-$3.139 &0.477 \\
5698279625842983680    &30.5893 &$-$1.236 $-$0.576 &0.211 \\
2026681443634299776    &30.2498 &$-$0.607 $-$0.394 &0.113 \\
5614343011257214720    &28.6367 &$+$0.514 $+$1.940 &0.332 \\
3360979795101021824    &27.9592 &$+$0.947 $-$0.516 &0.183 \\
5615665105256804736    &25.5830 &$+$0.839 $+$0.434 &0.175 \\
5677957524207024384    &22.0224 &$+$0.505 $+$0.344 &0.132 \\
OGLE BLG-ECL-73018     &14.6758 &$-$0.056 $-$0.273 &0.090 \\
TYC 1315-1096-1        &14.6163 &$+$0.143 $+$0.065 &0.051 \\
2059704828690939136    &12.9585 &$-$0.055 $-$0.096 &0.040 \\
4113264551321565952    &10.9226 &$-$0.192 $-$0.296 &0.153 \\
4359310510300687104    &9.9611  &$-$0.194 $+$0.156 &0.118 \\
4055988791630637312    &9.5297  &$+$0.157 $+$0.046 &0.081 \\
GALEX J013712.5-012958 &9.0019  &$-$0.007 $+$0.011 &0.007 \\
4658968939769587712    &7.5547  &$-$0.180 $-$0.033 &0.115 \\
OGLE BLG-ECL-126001    &7.3654  &$-$0.145 $+$0.005 &0.093 \\
GALEX J023702.5-353520 &6.6875  &$+$0.086 $+$0.109 &0.098 \\
OGLE BLG-ECL-37494     &5.0171  &$-$0.010 $-$0.030 &0.032

\end{longtable}
\end{center}
Potential close encounters are summarized in the following table supposing that $|v_r^{}|=50$ km/s.
\begin{center}
\begin{longtable}{|l|l|l|l|l|}
\caption{Closest stellar flybies (continuing Table 2).} 
\label{tab:flyby2} \\

\hline \multicolumn{1}{|c|}{\textbf{ID}} 
& \multicolumn{1}{c|}{$d_{\odot X}^{\mathrm{min}}$} 
& \multicolumn{1}{c|}{$d_{\odot X}^{\mathrm{min}}$}
& \multicolumn{1}{c|}{$d_{\odot X}^{\mathrm{min}}$}
& \multicolumn{1}{c|}{$t_{\odot X}^{\mathrm{min}}$} \\ 
\multicolumn{1}{|c|}{\textbf{}} 
& \multicolumn{1}{c|}{(pc)} 
& \multicolumn{1}{c|}{(ly)}
& \multicolumn{1}{c|}{(AU)}
& \multicolumn{1}{c|}{(kyr)} \\
\hline
\endfirsthead

\multicolumn{5}{c}%
{{\bfseries \tablename\ \thetable{} -- continued from previous page}} \\
\hline \multicolumn{1}{|c|}{\textbf{ID}} 
& \multicolumn{1}{c|}{$d_{\odot X}^{\mathrm{min}}$} 
& \multicolumn{1}{c|}{$d_{\odot X}^{\mathrm{min}}$}
& \multicolumn{1}{c|}{$d_{\odot X}^{\mathrm{min}}$}
& \multicolumn{1}{c|}{$d_{\odot X}^{\mathrm{min}}$} \\ \hline
\endhead

\hline \multicolumn{5}{|r|}{{Continued on next page}} \\ \hline
\endfoot

\hline
\endlastfoot

GALEX J013712.5-012958   &0.015 &0.050 &3147  &$\pm$2172.4 \\
UCAC4 638-030976         &0.027 &0.088 &5560  &$\pm$0285.2 \\
1940843120604424576      &0.036 &0.117 &7389  &$\pm$0555.7 \\
2059704828690939136      &0.062 &0.204 &12885 &$\pm$1509.1 \\
TYC 1315-1096-1          &0.070 &0.227 &14379 &$\pm$1337.9 \\
2026681443634299776      &0.075 &0.245 &15466 &$\pm$0646.5 \\
1835040758820804608      &0.106 &0.346 &21869 &$\pm$0493.9 \\
GALEX J002332.9+432030   &0.112 &0.366 &23164 &$\pm$0366.0 \\
5677957524207024384      &0.119 &0.390 &24638 &$\pm$0888.0 \\
OGLE BLG-ECL-73018       &0.123 &0.400 &25304 &$\pm$1332.5 \\
OGLE BLG-ECL-37494       &0.126 &0.411 &26002 &$\pm$3897.8 \\
3360979795101021824      &0.131 &0.427 &26979 &$\pm$0699.4 \\
5615665105256804736      &0.137 &0.446 &28224 &$\pm$0764.4 \\
5698279625842983680      &0.138 &0.451 &28499 &$\pm$0639.3 \\
5614676330771453440      &0.165 &0.538 &34004 &$\pm$0588.4 \\
4055988791630637312      &0.171 &0.557 &35229 &$\pm$2052.1 \\
UCAC4 362-032250         &0.190 &0.618 &39090 &$\pm$0507.7 \\
WISE J053516.80-750024.9 &0.193 &0.631 &39898 &$\pm$0078.0 \\
UCAC4 563-099325         &0.201 &0.656 &41504 &$\pm$0289.8 \\
4198241063381323264      &0.210 &0.683 &43224 &$\pm$0392.3 \\
UCAC4 563-093667         &0.210 &0.686 &43397 &$\pm$0341.5 \\
4158060785356113792      &0.213 &0.693 &43846 &$\pm$0588.2 \\
WISE J153541.67+222525.5 &0.223 &0.729 &46082 &$\pm$0215.3 \\
4143776239543076864      &0.231 &0.753 &47629 &$\pm$0534.1 \\
5614343011257214720      &0.232 &0.757 &47858 &$\pm$0682.9 \\
2031432910114027648      &0.233 &0.759 &47981 &$\pm$0476.7 \\
4359310510300687104      &0.238 &0.776 &49063 &$\pm$1963.2 \\
OGLE BLG-ECL-126001      &0.254 &0.827 &52301 &$\pm$2655.1 \\
5614623799042319360      &0.256 &0.837 &52903 &$\pm$0473.5 \\
UCAC4 767-032810         &0.259 &0.844 &53401 &$\pm$0245.2 \\
4113264551321565952      &0.280 &0.914 &57833 &$\pm$1790.4 \\
2027191994999462016      &0.285 &0.930 &58836 &$\pm$0613.8 \\
GALEX J023702.5-353520   &0.294 &0.960 &60711 &$\pm$2924.2 \\
5615515773536993920      &0.299 &0.975 &61650 &$\pm$0550.1 \\
4256872795844365568      &0.301 &0.982 &62128 &$\pm$0617.0 \\
4658968939769587712      &0.304 &0.992 &62704 &$\pm$2588.6 \\
UCAC4 486-052109         &0.309 &1.007 &63670 &$\pm$0328.7

\end{longtable}
\end{center}
When radial velocities will be measured, distances and times should be mutiplied by $50/|v_r^{}|$ approximately.

\section{Close flybies of massive bright stars}

In our previous paper [9], we examined an additional Nemesis candidate HD 107914 / HIP 60503 with small proper motions. It is a remarkable A7/8 giant, currently at $\approx$ 79.84 pc from the Sun. Here we remake calculations using new Gaia DR3 data.

\begin{center}
\begin{longtable}{|l|l|l|l|l|}
\caption{Astrometric data for massive bright stars.} \label{tab:giant} \\

\hline \multicolumn{1}{|c|}{\textbf{ID}} 
& \multicolumn{1}{c|}{$\pi_X^{}$} 
& \multicolumn{1}{c|}{proper motions}
& \multicolumn{1}{c|}{$v_t^{}$}
& \multicolumn{1}{c|}{$v_r^{}$} \\ 
& \multicolumn{1}{c|}{(mas)}
& \multicolumn{1}{c|}{(mas/yr)}
& \multicolumn{1}{c|}{(km/s)}
& \multicolumn{1}{c|}{(km/s)} \\ \hline
\endfirsthead

\multicolumn{5}{c}%
{{\bfseries \tablename\ \thetable{} -- continued from previous page}} \\
\hline \multicolumn{1}{|c|}{\textbf{ID}} 
& \multicolumn{1}{c|}{$\pi_X^{}$}
& \multicolumn{1}{c|}{prop. mot.} 
& \multicolumn{1}{c|}{$v_t^{}$}
& \multicolumn{1}{c|}{$v_r^{}$} \\ \hline 
\endhead

\hline \multicolumn{5}{|r|}{{Continued on next page}} \\ \hline
\endfoot

\hline
\endlastfoot

Algol     &36.27   &$+$2.99  $-$1.66  &0.447 &$+$4.0 \\
HD 107914 &12.5254 &$+$0.085 $+$0.966 &0.367 &$-$8.39 \\
HD 165704 &6.5505  &$-$0.112 $+$0.011 &0.081 &$-$17.77 \\
HD 2733   &4.2927  &$+$0.034 $+$0.009 &0.039 &$+$19.35

\end{longtable}
\end{center}

\smallskip An updated search indicates also a close encounter with B9-subgiant HD 165704, currently located at $\approx$ 152.7 pc from the Sun, 
and A7-subgiant HD 2733, located at $\approx$ 233.0 pc from the Sun.

\smallskip Algol ($\beta$ Persei) appears as well in the frames of section 3 with $|v_r^{}|=50$, though it's true radial velocity is only $\approx +4$ km/s.

\smallskip It gives us the following flyby data:
\begin{center}
\begin{longtable}{|l|l|l|l|}
\caption{Closest stellar flybies (continuing Tables 2 and 4).} \label{tab:giant2} \\

\hline \multicolumn{1}{|c|}{\textbf{ID}} 
& \multicolumn{1}{c|}{$d_{\odot X}^{\mathrm{min}}$} 
& \multicolumn{1}{c|}{$d_{\odot X}^{\mathrm{min}}$}
& \multicolumn{1}{c|}{$t_{\odot X}^{\mathrm{min}}$}\\ 
\multicolumn{1}{|c|}{\textbf{}} 
& \multicolumn{1}{c|}{(pc)} 
& \multicolumn{1}{c|}{(ly)}
& \multicolumn{1}{c|}{(kyr)} \\
\hline
\endfirsthead

\multicolumn{4}{c}%
{{\bfseries \tablename\ \thetable{} -- continued from previous page}} \\
\hline \multicolumn{1}{|c|}{\textbf{ID}} 
& \multicolumn{1}{c|}{$d_{\odot X}^{\mathrm{min}}$} 
& \multicolumn{1}{c|}{$d_{\odot X}^{\mathrm{min}}$}
& \multicolumn{1}{c|}{$t_{\odot X}^{\mathrm{min}}$} \\ \hline
\endhead

\hline \multicolumn{4}{|r|}{{Continued on next page}} \\ \hline
\endfoot

\hline
\endlastfoot

HD 2733   &0.468 &1.525 &$-$11771.5\\
HD 165704 &0.700 &2.282 &$+$8399.9 \\
Algol     &3.062 &9.986 &$-$6656.6 \\
HD 107914 &3.489 &11.380 &$+$9286.7

\end{longtable}
\end{center}

\section{Hypothetical pseudo-stream companion BD+60 596}

This chemically peculiar B1-star has simultaneously small radial velocity and proper motions. We would like to mention it despite a very high RUWE value in Gaia DR3 catalog.

\begin{center}
\begin{longtable}{|l|l|l|l|l|}
\caption{Astrometric data for BD+60 596.} \label{tab:BD} \\

\hline \multicolumn{1}{|c|}{\textbf{ID}} 
& \multicolumn{1}{c|}{$\pi_X^{}$} 
& \multicolumn{1}{c|}{proper motions}
& \multicolumn{1}{c|}{$v_t^{}$}
& \multicolumn{1}{c|}{$v_r^{}$} \\ 
& \multicolumn{1}{c|}{(mas)}
& \multicolumn{1}{c|}{(mas/yr)}
& \multicolumn{1}{c|}{(km/s)}
& \multicolumn{1}{c|}{(km/s)} \\ \hline
\endfirsthead

\multicolumn{5}{c}%
{{\bfseries \tablename\ \thetable{} -- continued from previous page}} \\
\hline \multicolumn{1}{|c|}{\textbf{ID}} 
& \multicolumn{1}{c|}{$\pi_X^{}$}
& \multicolumn{1}{c|}{prop. mot.} 
& \multicolumn{1}{c|}{$v_t^{}$}
& \multicolumn{1}{c|}{$v_r^{}$} \\ \hline 
\endhead

\hline \multicolumn{5}{|r|}{{Continued on next page}} \\ \hline
\endfoot

\hline
\endlastfoot

BD+60 596 &24.347 &$-$0.743 $+$0.696 &0.198 &$+$1.88262

\end{longtable}
\end{center}
It is one of the best candidates to be a pseudo-stream companion of the Sun.
\begin{center}
\begin{longtable}{|l|l|l|l|}
\caption{Closest stellar flybies (continuing Tables 2, 4 and 6).} \label{tab:BD2} \\

\hline \multicolumn{1}{|c|}{\textbf{ID}} 
& \multicolumn{1}{c|}{$d_{\odot X}^{\mathrm{min}}$} 
& \multicolumn{1}{c|}{$d_{\odot X}^{\mathrm{min}}$}
& \multicolumn{1}{c|}{$t_{\odot X}^{\mathrm{min}}$}\\ 
\multicolumn{1}{|c|}{\textbf{}} 
& \multicolumn{1}{c|}{(pc)} 
& \multicolumn{1}{c|}{(ly)}
& \multicolumn{1}{c|}{(kyr)} \\
\hline
\endfirsthead

\multicolumn{4}{c}%
{{\bfseries \tablename\ \thetable{} -- continued from previous page}} \\
\hline \multicolumn{1}{|c|}{\textbf{ID}} 
& \multicolumn{1}{c|}{$d_{\odot X}^{\mathrm{min}}$} 
& \multicolumn{1}{c|}{$d_{\odot X}^{\mathrm{min}}$}
& \multicolumn{1}{c|}{$t_{\odot X}^{\mathrm{min}}$} \\ \hline
\endhead

\hline \multicolumn{4}{|r|}{{Continued on next page}} \\ \hline
\endfoot

\hline
\endlastfoot

BD+60 596 &4.301 &14.027 &$-$21098.4 \\
\end{longtable}
\end{center}
So, about 21 Myr ago, BD+60 596 was close to the Sun. 

\section{Conclusion}

In this paper, we have established an extended list of stellar Nemesis candidates. However, radial velocities of many of them are still unknown. Also, more accurate measurements of proper motions are necessary too.

\smallskip We have used just a simple linear approximation without error estimations.

\smallskip This research has made use of the SIMBAD database, operated at CDS, Strasbourg, France.

\bibliographystyle{unsrt} 
\bibliography{flybies}

\bigskip
\begin{flushright}
Igor Potemine\\
Institut de Mathématiques\\
Universit\'e Paul Sabatier\\
118, route de Narbonne\\
31062 Toulouse (France)
\end{flushright}

\begin{flushright}
e-mail : igor.potemine@math.univ-toulouse.fr
\end{flushright}

\end{document}